\documentclass[iop]{emulateapj}
\usepackage{natbib}
\usepackage{color}
\slugcomment{}

\shorttitle{Ca and light elements abundances}
\shortauthors{Carretta et al.}

\begin{document}


\title{Calcium and light-elements abundance variations from high
resolution spectroscopy in globular clusters \altaffilmark{1}}


\author{Eugenio Carretta\altaffilmark{2}, Angela Bragaglia \altaffilmark{2},
Raffaele Gratton\altaffilmark{3}, Sara Lucatello\altaffilmark{3,4}, Michele
Bellazzini\altaffilmark{2}, Valentina D'Orazi\altaffilmark{3} }  

\altaffiltext{1}{Based on data collected at the European Southern Observatory,
Chile, programmes 072.D-507,073.D-0211,072.D-0742,077.D-0182}
\altaffiltext{2}{INAF, Osservatorio Astronomico di Bologna, via Ranzani 1,
        40127,  Bologna,  Italy. eugenio.carretta@oabo.inaf.it,
        angela.bragaglia@oabo.inaf.it, michele.bellazzini@oabo.inaf.it}
\altaffiltext{3}{INAF, Osservatorio Astronomico di Padova, vicolo
	dell'Osservatorio 5, 35122 Padova, Italy. raffaele.gratton@oapd.inaf.it,sara.lucatello@oapd.inaf.it, 
	valentina.dorazi@oapd.inaf.it}
\altaffiltext{4}{Excellence Cluster Universe, Technische Universit\"at 
         M\"unchen,Boltzmannstr. 2, D-85748, Garching, Germany}


\begin{abstract}

We use abundances of Ca, O, Na, Al from high resolution 
UVES spectra of 200 red giants in 17 globular clusters (GCs) to investigate  the
correlation found by \citet{lee09} between chemical enrichment from  SN II
and star-to-star variations in light elements in GC stars. We find that (i) the
[Ca/H] variations between first and second  generation stars are tiny in most
GCs ($\sim 0.02-0.03$~dex, comparable with typical  observational errors). In
addition, (ii) using a large sample of red giants in M~4 with  
abundances from 
UVES spectra from \citet{marino08}, we find that Ca and Fe abundances in
the two populations of Na-poor and Na-rich stars are identical.  These facts
suggest that the separation seen in color-magnitude diagrams using the  $U$\
band or $hk$\ index (as observed in NGC~1851 by \citealt{han09}) are not due  to
Ca variations. Small differences in [Ca/H] as associated to $hk$\ variations
might be due to a small systematic effect  in abundance analysis, because most
O-poor/Na-rich (He-rich) stars have slightly larger [Fe/H] (by 0.027~dex on
average, due to decreased H in the ratio) than first generation stars and are
then located at redder positions in the $V,hk$\ plane. While a few GCs (M~54,
$\omega$~Cen, M~22, maybe even NGC~1851) do actually show various degree of
metallicity spread, our findings eliminate the need of a close link between the
enrichment by core-collapse SNe with the mechanism responsible for the Na-O
anticorrelation.

\end{abstract}


\section{Introduction}

Stars of different stellar generations\footnote{We will use the words
stellar ``generation" and ``population" interchangeably, for the reasons
explained later.} are currently routinely found in all
Galactic globular clusters (GCs). A common misunderstanding is to identify {\it
multiple sequences} in the color-magnitude diagrams (CMDs) as the {\it only}
evidence of {\it multiple stellar populations} in GCs, while they are only the
tip of the iceberg, observable when differences in chemical composition and/or
age are  very large. However, all GCs display large spreads in
abundances of light elements. These are due to the presence of two generations
of stars, separated by a small difference in age, not enough to appear as a
smearing of the cluster turn-off, but clearly visible as a different chemical
signature. This difference can be uncovered by spectroscopy (see \citealt{gratton04}
for a recent review). Since its discovery by the Lick-Texas group (see the
early review by \citealt{kraft94}), the Na-O anticorrelation in GCs offers a powerful
instrument to resolve age differences as small as a few 10$^7$ yrs 
(should very massive stars be involved) by means
of a signal as large as 1 full dex in abundance. {\it Whenever the Na-O
anticorrelation is observed in a GC, multiple stellar generations are
present in the cluster}. 

The primordial P component of first generation stars, present in almost all 
Galactic GCs surveyed so far, has the high [O/Fe] and low [Na/Fe] ratios
typical of core-collapse supernovae (SNe II) nucleosynthesis, characteristic of
halo field stars with similar  metallicity. On the other hand,
\citet{carretta09a} showed that two other components of second generation stars
may be found: one dominant component has intermediate composition (I; observed
in $each$ GC), and another has extremely (E) modified composition, showing
large O-depletion and Na (and Al) enhancements, present only in the most
massive clusters\footnote{The operational distinction between I and E
components \citep{carretta09a} is not relevant here and we only consider the
total fraction (I+E) of second generation stars.}. This general pattern can be
qualitatively produced by proton-capture reactions in H-burning at high
temperature (\citealt{dd89,langer93}). The fact that
the same Na-O and Mg-Al  anticorrelations are found in unevolved stars in
several GCs (\citealt{gratton01,rc02,carretta04})
tells us that this composition is inherited from the ashes of a previous
generation of stars.

While the involved processes have been largely identified, it is still not
clear {\it where} these reactions occurred, the favorite sites being either
main-sequence core H-burning of fast rotating massive stars (FRMS, \citealt{decressin07}) or the hot bottom of the convective envelope in intermediate-mass
asymptotic giant branch (AGB) stars \citep{ventura01}. Whichever the sites
are, they can not synthesize iron or other elements heavier than Al. Indeed,
apart from a few notable exceptions ($\omega$~Cen, see reference in \citealt{gratton04}; M~22, \citealt{marino09,dacosta09}; M~54, \citealt{carretta10a}; Terzan~5, \citealt{ferraro09}; and possibly
NGC~1851,  \citealt{yg08}) the  level in [Fe/H] or $\alpha$ and Fe-group
elements is very constant in GCs \citep{carretta09b}.

However, a recent work by \citet[from now L09]{lee09} seems to challenge
this consolidated scenario, because in their Ca-$uvby$ survey they found a
spread in the photometric $hk$ index (including Ca {\sc ii} H and K
lines) for several GCs. They interpret this spread as due to Ca abundance
variations and claim that this $\alpha-$element was produced by SN II 
in a past phase of the
lifetime of GCs. In turn, these should be assumed to be initially much more
massive than at present, else the highly energetic SNe winds would have
been lost from the potential well. Moreover, they find that the Ca-strong group
in GCs with evidence of multiple stellar populations from spectroscopy is
associated to the population of stars with lower O and higher Na values, and
vice versa concerning the Ca-weak stars.

The need of massive proto-clusters is common to most scenarios of chemical
evolution of GCs, including one proposed by us \citep{carretta10b}. However
a correlation between Ca and light elements would imply that the second
generation stars were formed from material enriched not only by FRMS or AGB
stars, but also by SN II events, which is a substantial modification of the 
current scenario.

In the present Letter we exploit the large wealth of data from our ongoing
project ``Na-O anticorrelation and HB" \citep{carretta06} and from other
extensive spectroscopic surveys to compare [Ca/H] ratios obtained from high
resolution spectroscopy with the abundance distribution of O, Na, Al in a sample
of several hundred red giant branch (RGB) stars in 17 GCs. This allows to shed
light on the r\^ole of Ca in the chemical evolution of GCs; from our 
extensive spectroscopic database, we seek to confirm whether the spread in the
photometric $hk$ index is driven by a real spread in Ca abundance.

\section{Atmospheric parameters and abundance analysis}

The full description of the analysis is given in \citet{carretta09b}; here
we present only the abundances of Ca, which are based on a large number of lines
(typically 10-12 in the most metal-poor clusters like M~15
and almost 20 in metal-rich GCs like 47~Tuc). The [Ca/Fe] ratios with the number of
lines and the line-to-line r.m.s. scatter for each star are listed in Table
\ref{cah} (completely available on line only). In Table 2 we report
the number of stars in each cluster, the average [Ca/Fe] and [Ca/H] values and
the 1~$\sigma$ scatter of the mean. The rms scatters in [Ca/Fe] are smaller
than in [Ca/H] because the sensitivity of Ca and Fe lines to atmospheric
parameters is quite similar.
Five of the clusters listed in Table 1 are
included in the sample by L09, who claim significant spread in Ca for four of
them.

The values listed in Table~2 indicate that the mean level of Ca is highly
homogeneous in each GCs. The average r.m.s. star-to-star scatter for [Ca/Fe] is
0.03~dex from 17 GCs, to be compared with average scatter of 0.18 dex, 0.22 dex,
0.07 dex and 0.25 dex for O, Na, Mg, and Al respectively. Note that the mean
scatter in Mg, produced by core collapse SNe but also involved in proton-capture
reactions, is more than {\it twice} that of Ca, which does not participate in
any H-burning.

\section{Results}

Since we have only about 10 stars per cluster observed with UVES, we combined
the 17 GCs,  separating first generation (P) stars from second generation
(I+E) stars. For direct  comparison with L09, we used [Ca/H]
instead of [Ca/Fe], after normalizing the  individual values to the mean of each
cluster. In Figure 1 we show the histogram in [Ca/H]   for P and I+E
stars, with average values, errors, and rms.  About  two thirds of
stars belong to the second generation, and the average normalized [Ca/H] values 
do not differ from those of first generation (at about 1$\sigma$). A further
check comes from  the cumulative distributions, shown in Figure 2 (upper panel),
from which we excluded the 12 stars of NGC~2808, the GC in our sample showing
the most extreme variations in He (\citealt{piotto07,bra10}, and below in this
Letter). The two distributions cannot be distinguished.  Note that if we include
also NGC~2808, the probability of the Kolmogorov-Smirnov drops to 0.07, still
compatible with no significant difference, but possibly indicating some actual
variations in this particular cluster.
However, when we show in the
three lower panels of Figure 2  the cumulative distributions for
[O/Fe], [Na/Fe], and [Al/Fe], all elements involved in the (anti)correlations in
GCs, we clearly see a striking difference (compare e.g. to supplementary Fig.14
of L09).

Our sample of 200 stars comes from 17 different GCs, but there is already
available in literature a sample of similar size  for a single GC, M~4, for which
\citet{marino08} analyzed high resolution UVES spectra of 105 RGB stars.
This cluster is also in the L09 sample; from their
supplementary Fig.14 (panel i) we see that almost all stars defined Ca-strong
are Na-rich (in our interpretation, second generation I, E) and vice versa for
the Ca-weak. However, when we divide the stars in the Marino et al. sample in P
and I+E (using their same separation at [Na/Fe]=0.2) and compare the histograms
and the means of the two populations (see Figure 3, left panel), they look very
similar\footnote{As noticed by the referee, the same result holds using the
38 stars in NGC~6752 analyzed by \citet{yong05}. }. This is supported by 
the cumulative distributions (Figure 3, right
panel): they are indistinguishable, according to the KS test. A further check
was done considering only stars in the two extreme quartiles in
Na abundance: the average Ca values in the more Na-poor and in the more Na-rich stars
perfectly agree with the ones in Figure 3 for the I+E and P stars, respectively. 

{\em The stars
called Ca-weak  (Ca-strong) on the basis of their $hk$\ index do not seem to
have lower (higher) [Ca/H] values from direct analysis of high resolution
spectra; this  demonstrates that the differences in the $hk$\ index must be
produced by something other than Ca variations.}

It is important to stress that the observational effect at the basis of the L09
claim corresponds to 1$\sigma$ spreads in [Ca/H] of about 0.03 dex in
the most extreme  cases (NGC~1851, NGC~2808, excluding the special cases of
$\omega$ Cen and M~22 where the presence of a significant spread in iron has
already been established with high resolution spectroscopy) and of about 0.02 dex for
typical cases (like M~4, M~5, and NGC~6752, see Table 2)\footnote{We used their supplementary
Table 3 for the FWHM of the RGB of the GC we have in common, converted them back
to scatter in $hk$ assuming a Gaussian distribution, and further converted to
scatter in [Ca/H] and [Fe/H] using the sensitivity of Ca on the variation in
$hk$ (suppl. Sect. 4) and the relations in Sect. 3.}. 
In our view, these numbers call for two basic considerations:
\begin{enumerate} 
\item Such tiny spreads are similar in amplitude to several uncertainties that
are known to affect the abundance analysis (like e,g, variations in the He
content); they are usually neglected since the overall uncertainty on
single measures is typically larger than this. We do not exclude that there
might be dispersions at these very low levels, but presently they can not be
separated from the intrinsic errors.
\item even if we regard these very small variations as {\em real}, some basic
algebra indicates that, assuming a mass of $2\times 10^5~M_{\sun}$ and a
metallicity of $Z=10^{-3}$ for the gas cloud that gave origin to the cluster, a
single SN II event would be sufficient to produce the required amount of Ca (and
Fe).
\end{enumerate}

\section{The case for NGC~1851}

If the tiny variations detected in most GCs are not measuring intrinsic  scatter
in Ca, what might be the cause for the variations in the $hk$ index? L09
discussed several issues to show that the only way to affect the $hk$ index is a
change in Ca. Specifically, they made a strong case for NGC~1851, where the few
``Ca-strong" stars (with abundances from high resolution spectroscopy by \citet{yg08} are segregated to the reddest positions in the $V,hk$ CMD. In a
recent paper, \citet{han09} presented a split of the RGB in this cluster in
the $U,U-I$ plane, where the redder sequence is populated by Ca-strong stars.

We simply note the striking similarity between the CMD by Han et al. with the
$U,U-B$ CMD for M~4 by \citet{marino08}. In both cases, the RGB is clearly
split into two distinct sequences. However, the same dichotomy of RGB
sequences in M~4 is explained by \citet{marino08} as the N enhancement
affecting the $U-B$ colors (through the CN and NH molecular bands within the $U$
filter), so that Na-rich (and likely N rich) stars are separated from Na-Poor
(and N-poor) stars in the CMDs. Since we demonstrated in Figure~3 that in M~4
the [Ca/H] ratio is identical for the Na-poor and Na-rich stars, the same
conclusion might be valid for NGC~1851.

A further piece of the puzzle comes from \citet{carretta09a}, who confirmed
results by \citet{yong08} in NGC~6752: Na-poor stars  define a narrow N-poor
sequence all confined to the bluest part of the RGB in NGC~6752, whereas the
Na-rich (N-enhanced) stars are spread out to the red part of the RGB, reaching
large N abundances. Now, regardless of the nature of the polluters, the ejected
matter is always enriched in the main outcome of H-burning, He. \citet[in preparation]{bra10}
used a large sample (about 1400 RGB stars in 19 GCs) to
show that stars with higher He content should have on average slightly larger
[Fe/H] values the difference between the E and P stars being 0.027 dex in
[Fe/H], even if there is no difference in the overall metal content Z.

These findings suggest that
the correlation observed by L09 between the (small) dispersion in [Ca/H] and the
(large) spread in $hk$ may be related to variations in He and
light elements only, with no fresh production of Fe  and Ca. The sizable effect
on the Kolmogorov-Smirnov test of the inclusion of NGC~2808 in the sample
(Sect.~3) lends further support to this scenario, as this cluster seems to
display especially large variations in He abundance \citep{piotto07,bra10}.

Hence, while it is quite possible that in NGC~1851 some small intrinsic scatter
in iron does actually exist \citep{yg08}, our results do not support the idea that core collapse SNe
contributed to the enrichment of second generation stars in most GCs, and
indicate that the production of the proton-capture elements occurred in sites
not able to synthesize Ca and Fe-peak elements.

\section{Discussion and conclusions} 

Lee and coworkers devised a scenario for chemical enrichment in clusters with
multiple populations. It includes two-phases. First, core collapse SNe from
first generation stars inject in the intra-cluster medium material enriched with
Fe and $\alpha-$elements; at least part of this material is not lost by the GCs
owing to their larger mass at early epochs. Second generation stars form from
this enriched material, incorporating the ejecta of intermediate-mass AGB stars
with longer evolutionary times. The second generation stars are then produced
from matter enriched both in heavy and light elements.

However, in our opinion, this scenario has to face several problems:
\begin{itemize}
\item Since there is clear evidence of multiple populations from spectroscopy in
{\em all} GCs, except perhaps a few small ones  with only a handful of stars
analyzed (e.g. Pal 12, \citealt{cohen04}; Ter~7, \citealt{sbordone07}; both
belonging to the Sagittarius
dwarf galaxy), any proposed
scenario must account for the formation of the generality of GCs.
\item While a single core collapse SN might be enough to produce the claimed
scatter in [Ca/H], thousands (or even more) FRMS or massive AGB stars are
required to produce the observed pattern for elements involved in proton-capture
processes. The impact of these two mechanisms can not be the
same.
\item Core collapse SNe should produce a (variable) enhancement in the mean
abundances for other $\alpha-$elements. No significant variation has been
observed for Mg and Si (apart from those explained by the Mg-Al cycle; see Fig.
3 in \citealt{carretta09b}), or Ti \citep{gratton04}.
\item In GCs like M~4, a split in $hk$ of 45$\sigma_p(hk)$ (measurement errors
for the populations, L09, supplementary Fig. 10) between two
populations translates in virtually no difference in the [Ca/H] ratios obtained
from a homogeneous large sample of stars with high resolution spectra.
\item If FRMS are the polluters, the whole model would fail, since these stars
release the polluted matter in very short time, while still in MS. Hence, the
timescale is even shorter than that for production of core collapse SNe.
\item The model by L09, adapted to explain the double RGB seen in
NGC~1851, results in a number ratio of first-to second generation stars equal to
3 (75\%/25\%). From spectroscopy, on the other hand, it was found that the bulk
of stars in GCs belong to the second generation \citep{cohen02,carretta09b}, with a number ratio typically of about 0.5 (0.33/0.66). 
We stress that this feature is $not$ a prerogative of mono-metallic GCs,
the same value for the primordial fraction is also found in $\omega$ Cen and
M~54 \citep{carretta10a}, the two most massive clusters in
the Galaxy, both showing actual (large) dispersion in [Fe/H]: this finding
demonstrates that the relative fractions of first and second generation stars
are not related to the enrichment of heavy elements.
\item The $r-$process element europium, whose main production site is expected
to be  core-collapse SNe, is fairly constant within all the studied globular
clusters \citep{gratton04}.
\item Finally, one of the strongest objection to the L09 model comes from
the run of Al. With its yield extremely dependent on the neutron excess
(metallicity), any variations in the content of Fe (or heavy element produced in
SN II, like Ca) should be accompanied by clear changes in the primordial Al
abundance. As shown by \citet{carretta09b} this is not always seen, in
particular in clusters like M~4 where a bimodal distribution is claimed in the
$hk$ distribution. The Al distribution is far from being bimodal in M~4, on the
contrary [Al/Fe] is quite constant and does not vary between first and second
generation stars. This is clearly seen also in other GCs like NGC~6171, NGC~288,
and NGC~6838 from our own data \citep{carretta09b} as well in the larger
sample by \citet{marino08}. This constancy, due to the secondary
(metallicity-dependent) production of Al in SNe strongly precludes that any
significant difference in heavy metal-content is provided by SNe to stars in
different stellar generations in GCs.
\end{itemize}

We conclude that (i) the spreads in [Ca/H] are very small; (ii) different
distributions in O, Na, Al are clearly seen between first and second generation
stars in $all$ GCs analyzed so far, on the contrary no statistically significant
difference is found regarding Ca abundances from high resolution spectroscopy;
(iii) the small difference in Ca observed in some cases might be well explained
with systematic second-order effects due to the fact that second generation
stars are He-enhanced, and appear more metal-rich and hotter when analyzed with
high resolution spectroscopy.

At the moment, we cannot offer a viable alternative explanation for the 
dispersion in the $hk$ index observed by Lee and collaborators, although this
issue certainly requires further studies.

\acknowledgments
Partial funding come from PRIN MIUR 2007, PRIN INAF 2007, and the DFG cluster of
excellence ''Origin and Structure of the Universe''.

\clearpage

\begin{deluxetable}{lrccccc}
\tablecaption{[Ca/Fe] and [Ca/H] ratios for individual stars.
Complete table available on line only.\label{cah}}
\tablehead{\colhead{GC ID}&
\colhead{star}&
\colhead{nr.}&
\colhead{[Ca/Fe]}&
\colhead{$\sigma$}&
\colhead{[Ca/H]}&
\colhead{flag PIE}\\
}
\startdata
NGG0104 &  5270 & 15 & 0.299 & 0.150 & $-$0.473 & 2 \\
NGG0104 & 12272 & 14 & 0.310 & 0.107 & $-$0.437 & 2 \\
NGG0104 & 13795 & 11 & 0.345 & 0.105 & $-$0.487 & 1 \\
NGG0104 & 14583 & 14 & 0.309 & 0.110 & $-$0.375 & 2 \\
NGG0104 & 17657 & 11 & 0.298 & 0.127 & $-$0.546 & 3 \\
NGG0104 & 18623 & 19 & 0.322 & 0.213 & $-$0.451 & 2 \\
\enddata
\tablenotetext{a}{nr is the number of Ca lines in each star.}
\tablenotetext{b}{flag is 1,2,and 3 for the primordial P, intermediate I and
extreme E component, respectively; flag 0 means that either Na or O was missing
and the star could be no assigned to a component.}
\end{deluxetable}

\clearpage

\begin{deluxetable}{lrcccccc}
\tabletypesize{\scriptsize}
\tablecaption{Mean [Ca/Fe], [Ca/H] ratios and internal errors. \label{meanca}}
\tablehead{\colhead{GC ID}&
\colhead{nr}&
\colhead{[Ca/Fe]}&
\colhead{$\sigma$}&
\colhead{err1}&
\colhead{err2}&
\colhead{[Ca/H]}&
\colhead{$\sigma$}\\
}
\startdata
NGC~0104 & 11 & +0.31 & 0.01 & 0.03 & 0.03 & $-$0.45 & 0.06\\
NGC~0288 & 10 & +0.41 & 0.02 & 0.02 & 0.03 & $-$0.90 & 0.05\\
NGC~1904 & 10 & +0.28 & 0.01 & 0.02 & 0.02 & $-$1.30 & 0.04\\
NGC~2808 & 12 & +0.34 & 0.02 & 0.03 & 0.03 & $-$0.81 & 0.07\\
NGC~3201 & 13 & +0.30 & 0.03 & 0.02 & 0.02 & $-$1.21 & 0.07\\
NGC~4590 & 13 & +0.26 & 0.03 & 0.03 & 0.03 & $-$2.00 & 0.04\\
NGC~5904 & 14 & +0.38 & 0.02 & 0.02 & 0.03 & $-$0.96 & 0.05\\
NGC~6121 & 14 & +0.42 & 0.03 & 0.02 & 0.03 & $-$0.75 & 0.07\\
NGC~6171 &  5 & +0.41 & 0.02 & 0.03 & 0.03 & $-$0.63 & 0.08\\
NGC~6218 & 11 & +0.42 & 0.03 & 0.03 & 0.03 & $-$0.91 & 0.04\\
NGC~6254 & 14 & +0.34 & 0.04 & 0.02 & 0.02 & $-$1.23 & 0.07\\
NGC~6397 & 13 & +0.28 & 0.03 & 0.03 & 0.03 & $-$1.71 & 0.05\\
NGC~6752 & 14 & +0.40 & 0.03 & 0.03 & 0.03 & $-$1.16 & 0.04\\
NGC~6809 & 14 & +0.36 & 0.04 & 0.02 & 0.02 & $-$1.58 & 0.07\\
NGC~6838 & 12 & +0.31 & 0.06 & 0.03 & 0.03 & $-$0.53 & 0.07\\
NGC~7078 & 13 & +0.25 & 0.05 & 0.03 & 0.03 & $-$2.07 & 0.10\\
NGC~7099 & 10 & +0.31 & 0.04 & 0.03 & 0.03 & $-$2.03 & 0.05\\
\enddata
\tablenotetext{a}{nr is the number of stars in each cluster.}
\tablenotetext{b}{err1 for [Ca/Fe] account for internal errors in T$_{\rm eff}$, $v_t$
and $EW$, whereas err2 includes all sources of uncertainties (see Carretta et al. 2009b).}
\end{deluxetable} 

\clearpage

\begin{figure}
\centering
\includegraphics{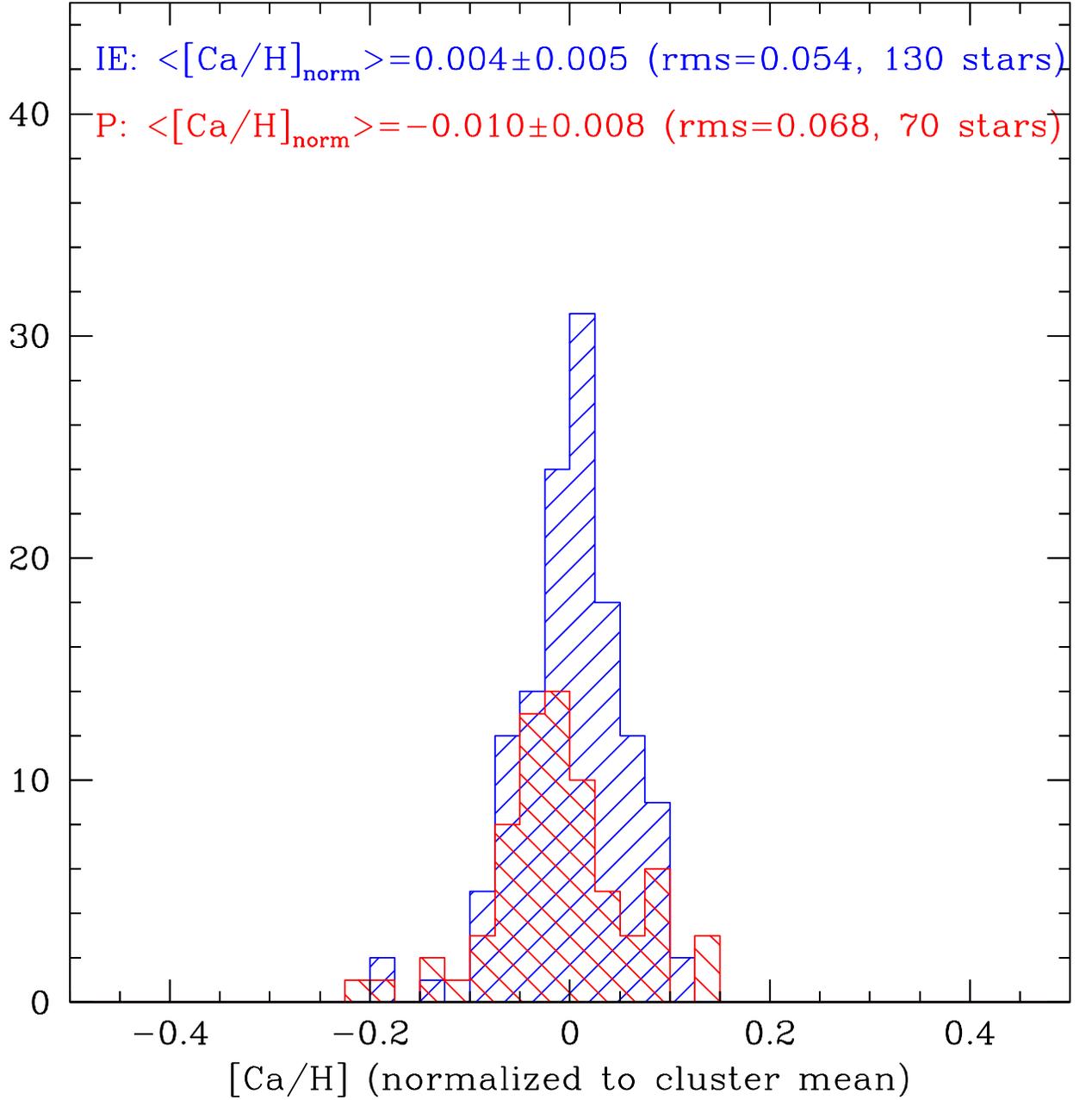}
\caption{Distribution of [Ca/H] ratios from high resolution UVES spectra for
202 RGB stars in 17 globular clusters. The cross-hatched red area is the
histogram for stars of the primordial P component, while the blue hatched area
indicates stars of the second generation (I+E stellar component, see text.)}
\label{fig1}
\end{figure}

\clearpage

\begin{figure}
\centering
\includegraphics{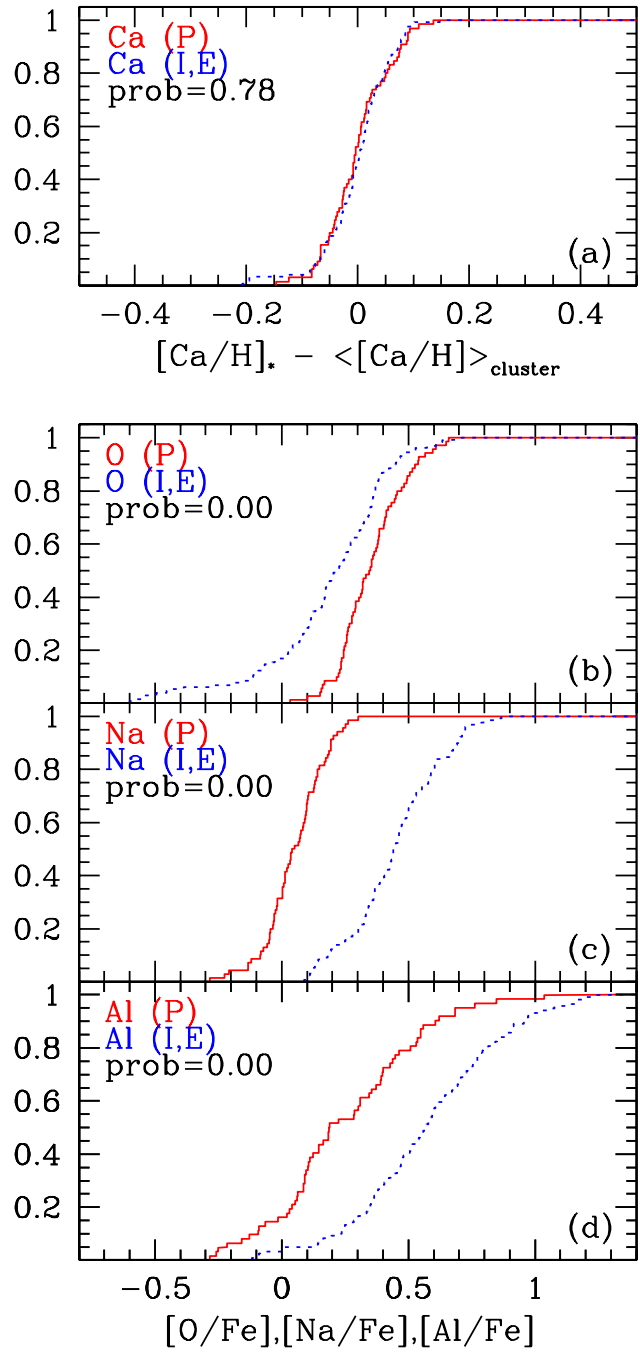}
\caption{Cumulative distributions for stars of the P component (solid red lines)
and for second generation stars (I+E components, dotted blue lines). From top
to bottom the distribution of abundances of [Ca/H], [O/Fe], [Na/Fe] and 
[Al/Fe] are plotted. In the top panel NGC~2808 is excluded (see text); if included the
KS probability would become 0.07 .}
\label{fig2}
\end{figure}

\clearpage

\begin{figure}
\centering
\includegraphics{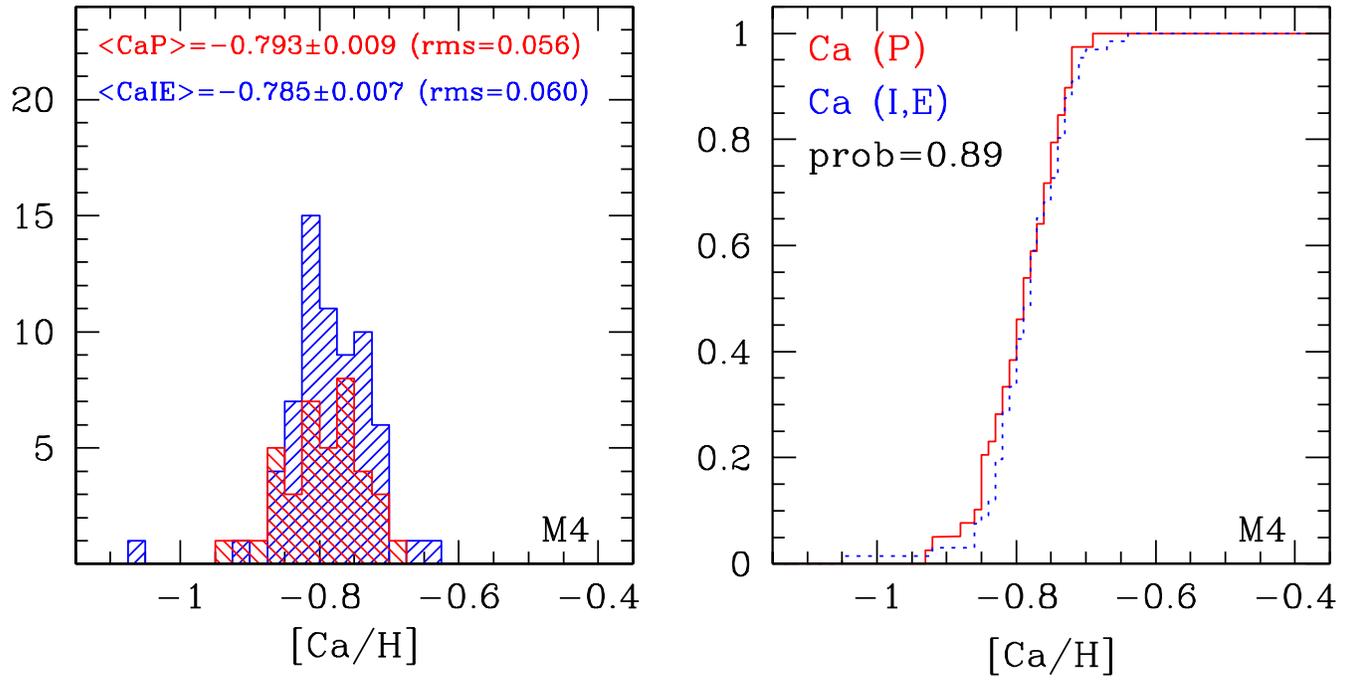}
\caption{Distribution of [Ca/H] ratios in M~4 (from UVES spectra by \citealt{marino08}; left panel), and cumulative distributions for stars more Na-rich
(blue) and
more Na-poor (red) than [Na/Fe]=$+0.2$ dex, that we identify with the I+E and P
components, respectively.}
\label{fig3}
\end{figure}

\end{document}